\documentclass[%
 reprint,
bibnotes,
 amsmath,amssymb,
 aps
]{revtex4-1}

\usepackage{graphicx}
\usepackage{dcolumn}
\usepackage{xr}
\usepackage{bm}
\usepackage{hyperref}
\usepackage{color,soul}

\makeatletter
\newcommand*{\addFileDependency}[1]{
  \typeout{(#1)}
  \@addtofilelist{#1}
  \IfFileExists{#1}{}{\typeout{No file #1.}}
}
\makeatother

\newcommand*{\myexternaldocument}[1]{%
    \externaldocument{#1}%
    \addFileDependency{#1.tex}%
    \addFileDependency{#1.aux}%
}

\myexternaldocument{SI}

\begin{document}


\title{Quantum optomechanics in tripartite systems}

\author{Ryan O. Behunin}
\affiliation{Department of Applied Physics and Materials Science, Northern Arizona University, Flagstaff, AZ 86011}
\affiliation{Center for Materials Interfaces in Research and Applications (!`MIRA!), Northern Arizona University, Flagstaff, Arizona 86011}
\author{Peter T. Rakich}
\affiliation{Department of Applied Physics, Yale University, New Haven, Connecticut 06520}

\date{\today}

\begin{abstract}
Owing to their long-lifetimes at cryogenic temperatures, mechanical oscillators are recognized as an attractive resource for quantum information science and as a testbed for fundamental physics. Key to these applications is the ability to prepare, manipulate and measure quantum states of mechanical motion.  Through an exact formal solution to the Schrodinger equation, we show how tripartite optomechanical interactions, involving the mutual coupling between two distinct optical modes and an acoustic resonance enables quantum states of mechanical oscillators to be synthesized and interrogated.
\end{abstract}
     
\maketitle

 
New methods to prepare and interrogate nonclassical states of mechanical oscillators could enable novel quantum technologies as well as the exploration of fundamental physics \cite{Marshall2003,Goryachev2012,Goryachev2013,Galliou2013a,Lo2016,Tobar2017,Renninger2018,Kharel2018,Eichenfield2009,Eichenfield2009a,weaver2017,campbell2021}. If the astounding lifetimes exhibited by phonons at cryogenic temperatures  \cite{Goryachev2012,maccabe2020nano} can be translated to quantum coherence times, phononic systems could form the basis for high-dimensional quantum memories \cite{pechal2019}. 
In addition, the ability to interrogate and manipulate these acoustic modes using superconducting qubits \cite{Chu2017,Chu2018,wollack2022,manenti2017}, electrical signals \cite{Goryachev2012,Goryachev2013,Galliou2013a}, or telecommunications wavelengths of light \cite{Renninger2018,Kharel2018,maccabe2020nano} make them compelling candidates for quantum repeaters \cite{pechal2019} and high-fidelity quantum state-transfer \cite{weaver2017,wollack2022,reed2017faithful}. Whereas, mechanical oscillators with large effective mass may shed light on the quantum-to-classical transition \cite{Ghirardi1990,Percival2006}, the nature of dark matter \cite{manley2020,campbell2021}, and the impacts of gravity on decoherence \cite{Marshall2003,diosi1989models,penrose2014gravitization}.

Generation, control, and measurement of quantum states of mechanical oscillators has recently been explored in a variety electromechanical and optomechanical systems \cite{o2010quantum,palomaki2013entangling,Aref2015,nielsen2017multimode,hong2017hanbury,Chu2017,Chu2018,sletten2019,satzinger2018quantum,wollack2022}.  Within circuit systems, nonlinearity provided by a superconducting qubit has enabled quantum state preparation and readout in the mechanical domain \cite{o2010quantum,Aref2015,Chu2017,Chu2018,sletten2019,satzinger2018quantum,wollack2022}. Canonical cavity optomechanical interactions, that utilize nonlinear coupling between a single electromagnetic mode and a single mechanical mode (i.e., bi-partite system), permit an array of state preparation, control and readout functionalities \cite{aspelmeyer2014cavity,mancini1997,Bose1997,hong2017hanbury,palomaki2013entangling}. By detuning a strong coherent drive from resonance, a linearized optomechanical coupling can be realized, enabling coherent state swaps between the mechanical and optical domains, ground-state cooling, entanglement generation, two-mode squeezing, and when combined with photon number measurements, the synethsis of single-phonon Fock states \cite{aspelmeyer2014cavity,palomaki2013entangling,hong2017hanbury}. 

Looking beyond these demonstrations, it is challenging to access more exotic quantum states using conventional bipartite cavity optomechanical systems. While it is possible to create multi-component cat states, macroscopically distinguishable superpositions, and phonon-photon entanglement if one can reach the ultrastrong coupling regime, this regime requires coupling rates on par with the phonon frequency \cite{Bose1997}. 
Moreover, relatively weak optomechanical nonlinearities make this regime difficult to access with GHz-frequency phonons, which offer long coherence times at cryogenic temperatures. Alternatively, high frequency phonons can be accessed using a tripartite system consisting of a single phonon mode that mediates coupling between two optical modes \cite{Kharel2018,Renninger2018}. Moreover, the distinct structure of the tripartite system may offer some unique advantages as we consider new strategies to generate and detect exotic quantum states with mechanical systems. 

Here, we show that the nonlinear quantum dynamics of tripartite optomechanical systems can enable the preparation of highly nonclassical phononic states. Considering a triply resonant system, we derive a formal solution for the exact time-evolution of the total system wavefunction, enabling analytical and numerical calculations for the quantum state dynamics. Our results show that experimentally accessible initial states (e.g., prepared using a coherent classical drive) evolve into wavefunctions exhibiting entanglement between optical and mechanical degrees of freedom. Leveraging this entanglement, we show that conditional measurements on the optical modes of the system, such as homodyne detection and/or photon counting, can project the mechanical oscillator into highly nonclassical states that depend sensitively on the initial system wavefunction. By simulating the system evolution including the effects of decoherence, we identify regimes where quantum states can be robustly synthesized. Moreover, in the presence of a classical coherent drive, we show that the phonon's reduced density matrix exhibits nonclassicality even without state-collapsing conditional measurements. We also illustrate how $\pi/2$- and $\pi$-pulses can be used to entangle optical and mechanical modes, or transfer quantum states between the optical and mechanical domains. 

While closely related, this tripartite system has important features not present in the canonical cavity optomechanical interaction that afford unique quantum dynamics. First, access to two optical modes greatly expands the number and complexity of the phonon states that can be heralded, where projective measurements produce families of phonon states parameterized by two sets of observables. 
Second, photon number measurements can herald highly nonclassical phonon states, in contrast with standard resonant cavity optomechanical interactions, even in systems with weak coupling \cite{Bose1997}. Third, for telecommunications wavelengths of light the relevant phonon frequencies are of order $\sim$10 GHz, enabling ground state cooling with standard cryogenics. Put together, these results reveal an unexplored regime of nonlinear quantum dynamics in systems spanning from chip-scale optomechanical devices \cite{Eichenfield2009,Eichenfield2009a} to bulk crystals \cite{Renninger2018,Kharel2018}. 

{\it Quantum Dynamics}: To illustrate how quantum state generation can be accomplished using multi-mode optomechanical coupling, we explore the dynamics of a system described by the Hamiltonian $H = H_0 +H_{int}$, 
\begin{align}
\label{H}
 H_0   & =  \hbar \omega_p a^\dag_p a_p + \hbar \omega_S a^\dag_S a_S   + \hbar \Omega b^\dag b, \\
     \nonumber
 H_{int}      & =    \hbar g( a_p a^\dag_S b^\dag + a_p^\dag a_S b).
\end{align}
Here, $a_p$, $a_S$ and $b$, are the annihilation operators of the pump, Stokes, and phonon modes, with angular frequencies $\omega_p$, $ \omega_S$, and $\Omega$, respectively. This interaction Hamiltonian, $H_{int}$, describes phonon-mediated coupling between these two electromagnetic modes. Throughout, we assume that our system satisfies the condition $\omega_p = \omega_S + \Omega$, necessary for the phonon mode to mediate resonant coupling between the photon modes (i.e., inter-modal scattering). 
Systems that are well described by this Hamiltonian typically utilize a high-frequency elastic wave to mediate resonant inter-modal scattering (e.g., through Brillouin interactions \cite{kharel2019high}), with couplings ($g$) that can be produced by electrostriction or radiation pressure \cite{Rakich2012,Kharel2016}. In the analysis that follows, we consider the dynamics of this system for times that are much shorter than the decoherence time of our phonon mode \cite{Chu2017,satzinger2018quantum}, permitting us to neglect the effects of phonon decoherence.

Neglecting decoherence, application of the time evolution operator to the initial wavefunction gives the quantum dynamics of this system in terms of the time-dependent wavefunction, given by the formal solution to the Schrodinger equation 
$
 | \psi(t) \rangle = \exp\{-iH_0 t/\hbar\}\exp\{-iH_{int} t/\hbar\} | \psi(0) \rangle.
$
Because $\omega_p = \omega_S + \Omega$, $H_0$ and $H_{int}$ commute, permitting the time-evolution operator to be factorized. While the operator $\exp \{ -i H_{int} t/\hbar\}$ is an exponent of non-commuting operators, a symmetry of the system provides a path to a formal analytical solution: for a Fock state, the total number of phonons and pump photons $n_p+n_b$ is conserved, reducing the Hilbert space to a compact $(n_p+n_b+1)$-dimensional subspace. Within this compact Hilbert space, $H_{int}$ can be diagonalized, where the Hamiltonian given by Eq. \eqref{H} is formally equivalent to a Jaynes-Cummings model describing the interaction between a bosonic mode and a spin-$(n_p+n_b)/2$ system (see Supplementary Information \ref{SI:FormalEquiv}) \cite{jaynes1963comparison}. 

For the initial state $|n,m,0\rangle \equiv | n \rangle_p \otimes | m \rangle_{S} \otimes | 0 \rangle_{ph}$, where the pump, Stokes and phonon modes respectively have $n$, $m$ and $0$ quanta and using Eq. \eqref{H}, the time-dependent wavefunction $| \psi_{nm0}(t) \rangle$ in the interaction picture is generally represented by 
\begin{align}
\label{WFnm}
 | \psi_{nm0}(t) \rangle=\sum_{k=0}^n A_{n,m,k}(t) |n-k,m+k,k \rangle. 
\end{align}
Truncation of the sum over $k$ at $n$ is a consequence of the compact nature of the Hilbert space for Fock state evolution. 
Inserting $| \psi_{nm0}(t) \rangle$ into the Schrodinger equation yields a linear matrix differential equation for the complex probability amplitudes $A_{n,m,k}(t)$ given by 
\begin{eqnarray}
\label{probAmps}
\dot{\vec{A}}_{n,m}
= -i {\bf M}_{nm} \cdot \vec{A}_{n,m}.
\end{eqnarray}
Here, $\vec{A}_{n,m}$ is a column vector of the probability amplitudes $
\vec{A}_{n,m} = 
(
A_{n,m,n},
A_{n,m,n-1},
\hdots 
,A_{n,m,0} 
)^T$, and
${\bf M}_{nm}$ is the symmetric matrix
\begin{eqnarray}
\label{}
{\bf M}_{nm}
= 
\begin{pmatrix}
0 			& \Lambda^{nm}_{n} 	&         0 				& 0 			& \hdots \\ 
\Lambda^{nm}_{n}	& 0 				& \Lambda^{nm}_{n-1}	& 0 			& 	     \\
0 			&  \Lambda^{nm}_{n-1} 	& 0 					& 0 			& 	     \\
0			&		0			&\ddots&              		& \vdots \\ 
\vdots	 	&					&					& 0 			&\Lambda^{nm}_{1} \\
	 		&					&	\hdots			& \Lambda^{nm}_{1}	& 0 \\
\end{pmatrix}
\end{eqnarray}
with matrix elements given by $\Lambda^{nm}_{k} =  g  \sqrt{n-k+1}\sqrt{m+k}\sqrt{k}$. 
The solution to Eq. \eqref{probAmps} can be obtained by diagonalizing the matrix ${\bf M}_{nm}$, yielding
\begin{eqnarray}
\label{probAmpsSolution}
\vec{A}_{n,m}(t) = {\bf V}_{nm} \cdot e^{-i {\bf \Omega}_{nm} t} \cdot {\bf V}_{nm}^\dag \cdot \vec{A}_{n,m}(0)
\end{eqnarray}
where ${\bf V}_{nm}$ is a unitary matrix diagonalizing ${\bf M}_{nm}$, and the diagonal matrix of eigenvalues ${\bf \Omega}_{nm} = {\bf V}^\dag_{nm}\cdot {\bf M}_{nm}\cdot {\bf V}_{nm}$ \cite{arfken1999mathematical} (see Supplementary Information Sec. \ref{DiagonalM}). 

Focusing on initial states that can be prepared in the laboratory using classical light sources, photon squeezing, or single photon emitters, we calculate the system wavefunction. With the phonon cooled to the ground state, we consider initial wavefunctions given by 
\begin{eqnarray}
\label{psi_0}
|\psi(0)\rangle = \sum_{n=0}^\infty \sum_{m=0}^\infty \mathcal{P}_n \mathcal{S}_m |n,m,0\rangle
\end{eqnarray}
where $\mathcal{P}_n$ ($\mathcal{S}_m$) is the probability amplitude for the pump (Stokes) mode to be found initially in the $n$th ($m$th) Fock state. Using Eqs. \eqref{probAmpsSolution} \& \eqref{psi_0}, the time-dependent wavefunction is given by  
\begin{align}
\label{WF}
& | \psi(t) \rangle  = \sum_{n,m=0}^\infty \sum_{k=0}^n \mathcal{P}_n \mathcal{S}_m A_{n,m,k}(t) |n-k,m+k,k \rangle. \end{align}
with time-dependent coefficients $A_{n,m,k}(t)$ given by
\begin{align}
\label{WFcoeffs}
 & A_{n,m,k}(t)=\hat{k}\cdot {\bf V}_{nm} \cdot e^{-i {\bf \Omega}_{nm} t} \cdot {\bf V}_{nm}^\dag \cdot \vec{A}_{n,m}(0)
\end{align}
where $A_{n,m,k}(0) = \delta_{k0}$ and  $\hat{k}$ is a unit vector of dimension $n+1$ given by $\hat{k} = (\delta_{n,k},\delta_{n-1,k}, \hdots, \delta_{k0})$ (see Supplementary Information Sec. \ref{sec:probabilityAmplitudes} for a list of $A_{n,m,k}(t)$ for $n=0$ to $2$). For optical states prepared with lasers $\mathcal{P}_n = \alpha_p^n \exp\{-|\alpha_p|^2/2\}/\sqrt{n!}$, describing a coherent state of amplitude $\alpha_p$, or for single photon emitters with $\mathcal{P}_n = \delta_{n1}$---definitions that also apply to the Stokes probability amplitudes $\mathcal{S}_m$. For example, when the system is prepared with a single pump photon, the wavefunction is given by
\begin{align}
\label{WF1p}
 | \psi(t) \rangle = \sum_{m=0}^\infty \mathcal{S}_m &  
 \big[ \cos(g\sqrt{m+1} t) |1,m,0\rangle \nonumber \\
&  - i \sin(g\sqrt{m+1} t) |0,m+1,1\rangle\big]. 
\end{align}
In this case, the total system dynamics is analogous to the Jaynes-Cummings model, exhibiting Rabi-oscillations and coherent energy exchange between the three modes \cite{jaynes1963comparison,shore1993jaynes} (see Supplementary Information Sec. \ref{SI:FormalEquiv}).

Even when the input optical fields are classical, quantum states can be generated. To illustrate, consider weak coherent states in the pump and Stokes modes, i.e., $|\alpha_p|, |\alpha_S| \ll 1$, where the initial photonic states are well approximated by a superposition of the vacuum and first excited states ($|\alpha_p\rangle \approx |0\rangle + \alpha_p |1\rangle$). In this limit the wavefunction is given by 
\begin{align}
\label{WFsa}
 | \psi(t) \rangle \approx & 
 |0,0,0\rangle +
 \alpha_S e^{-i \omega_S t}  |0,1,0\rangle  +  \\
& + \alpha_p e^{-i \omega_p t} (\cos(g t)  |1,0,0\rangle -i \sin(gt)  |0,1,1\rangle ) \nonumber
\end{align}
 to first order in $\alpha_P$ and $\alpha_S$, exhibiting quantum entanglement. For larger amplitudes, one must include more terms in the series representation of the wavefunction given in Eq. \eqref{WF}.  
Nevertheless, this example clearly illustrates how classical coherent states can be used to generate quantum states with resonant three-mode coupling (or with resonant coupling in this tripartite system). 
 
\begin{figure}[h]
\vspace{-0.0 in}
\includegraphics[width=8cm]{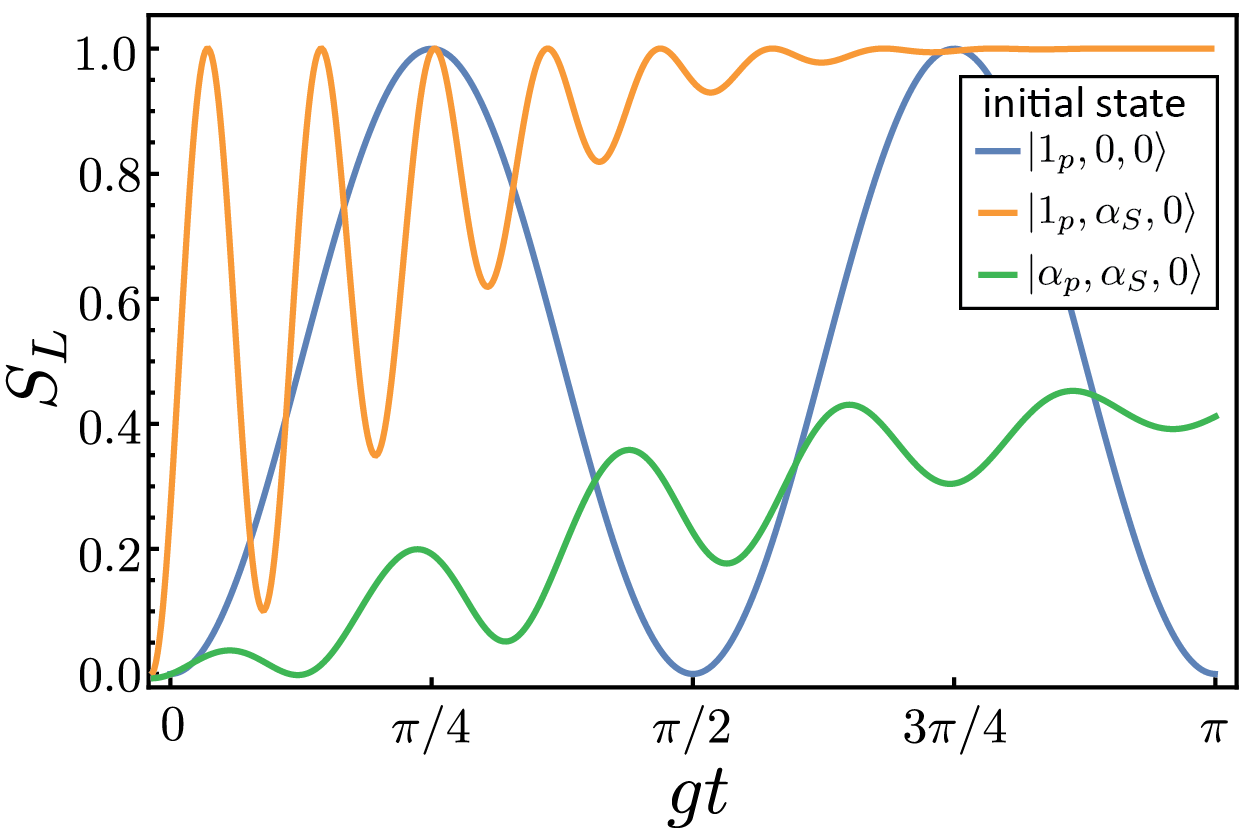}
\begin{center}
\vspace{-0.25 in}
\caption{Phonon linear entropy for experimentally accessible initial states. Coherent state amplitudes $\alpha_p = 1.0$ and $\alpha_S = 4.6$. For the single pump photon initial states, the linear entropy is scaled by a factor $2$ so that $S_L$ varies between 0 and 1. $S_L =1$ indicates maximal phonon-photon entanglement.}
\vspace{-0.45 in}
\label{S_L}
\end{center}
\end{figure}

These results show that resonant inter-modal coupling coupling can produce entanglement between the mechanical oscillator and the optical modes, even for classically prepared initial states (e.g., Eq. \eqref{WFsa}).  
Looking beyond this analytical example of Eq. \eqref{WFsa}, we can use the linear entropy $S_L(t) \equiv 1- {\rm Tr} \{ \hat{\rho}_{ph}^2(t) \}$, where $\hat{\rho}_{ph}(t) = {\rm Tr}_{p,S} \{ |\psi(t) \rangle \langle \psi(t) | \}$ is the reduced density matrix for the phonon mode, to analyze the degree of entanglement produced by more complex quantum states, evaluated through the numerical evaluation of Eq. \eqref{WF}.
Figure \ref{S_L} quantifies the entanglement between the phonon mode and the optical fields for a selection of initial states, exhibiting temporal oscillations determined by the coupling rate $g$, the initial state, and the eigenvalues ${\bf \Omega}_{nm}$.
Using Eq. \eqref{WF} summed to $>99\%$ convergence, Fig. \ref{S_L} shows that phonon-photon entanglement persists for larger coherent state amplitudes.

{\it Preparing and measuring quantum states of a mechanical mode}: Leveraging phonon-photon entanglement shown by Eq. \eqref{WF} and Fig. \ref{S_L}, the phonon mode can be projected into a large variety of highly nonclassical states through conditional measurements of the optical fields. By applying the projective operator $\hat{P} = | \varphi_{p}, \varphi_{S} \rangle \langle \varphi_{p}, \varphi_{S} |$ to $|\psi(t)\rangle$, we obtain the phonon wavefunction $|\psi_{ph}[ \varphi_{p}, \varphi_{S},t) \rangle \equiv  \langle \varphi_{p}, \varphi_{S}| \psi(t) \rangle$ resulting from the conditional measurement of the optical fields with measurement outcomes given by $\varphi_p$ and $\varphi_S$ for the pump and Stokes modes respectively. For example, for homodyne detection $\varphi_{p}$ represents the measured complex coherent state amplitude of the pump mode, or a photon number resolving measurement $\varphi_{p}$ indicates the measured number of pump photons.
\begin{figure}[]
\includegraphics[width=0.45\textwidth]{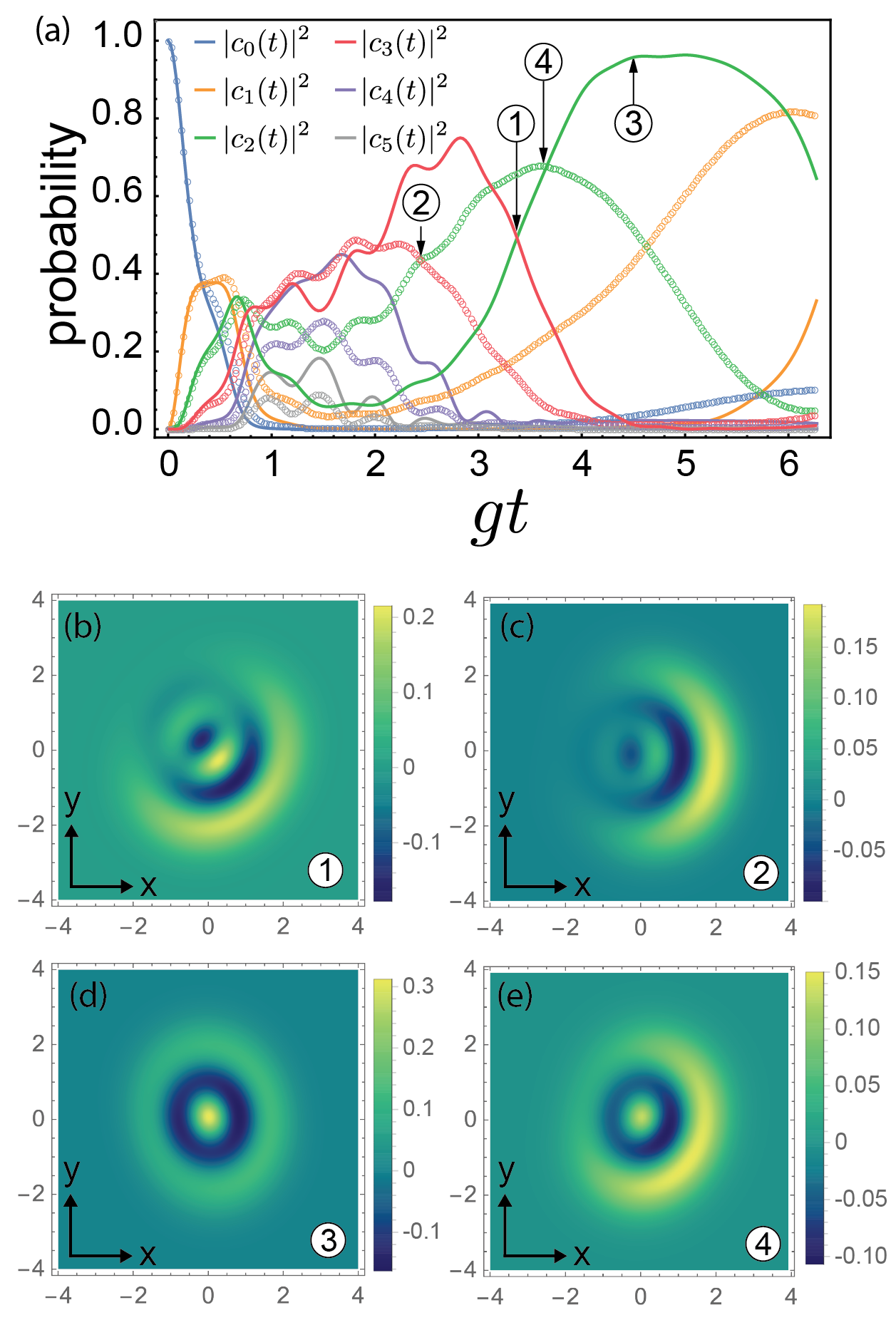}
\begin{center}
\vspace{-0.2 in}
\caption{Conditional quantum state preparation through homodyne measurements of the optical field. The initial state is $|\alpha_p, \alpha_S,0\rangle$ and the optical fields are measured in state $|\bar{\alpha}_p,\bar{\alpha}_S\rangle$ at a later time. For $\alpha_p = 0.74, \ \alpha_S= 5.6, \ \bar{\alpha}_p = 0.01, \  \bar{\alpha}_S= 5.65 \exp\{i 0.3 \pi\}$. (a) Phonon Fock state probabilities calculated from Eq. \eqref{phPA} (solid lines) and using a numerical master equation solver with damping for both pump and Stokes modes of $\gamma = 0.25 g$ (open circles). (b)-(e) Wigner functions for targeted phonon states with and without optical losses. States approximating $|\psi_{ph}\rangle \approx (|2\rangle + |3\rangle \exp\{i\phi\})/\sqrt{2})$ shown in (b) $gt = 3.37$, denoted by $\textcircled{1}$ (lossless) and (c) $gt = 2.63$, denoted by $\textcircled{2}$ (optical losses). States approximating $|\psi_{ph}\rangle \approx |2\rangle$ shown in (d) $gt = 4.5$, denoted by $\textcircled{3}$ (lossless), and (e) $gt = 3.63$, denoted by $\textcircled{4}$ (optical losses). Fidelities for state preparation with losses are 72\% for (c) and 67\% for (e). }
\vspace{-0.5 in}
\label{fig:wigHomodyne}
\end{center}
\end{figure}

Using homodyne detection, a projective measurement of the amplitude and phase of the optical modes in the interaction picture collapses the phonon wavefunction into the superposition of Fock states given by 
$
 | \psi_{ph}[\bar{\alpha}_p,\bar{\alpha}_S,t) \rangle =  \sum_{k=0}^\infty c_k(t) |k \rangle.
$
Here, the unnormalized probability amplitudes $c_k(t)$ are given by 
\begin{align}
\label{phPA}
 c_k(t) =  \sum_{n=k}^\infty& \sum_{m=0}^\infty  \frac{\alpha_p^n  (\bar{\alpha}_p^*)^{n-k}\alpha_S^m (\bar{\alpha}_S^*)^{m+k} }{\sqrt{n!}\sqrt{(n-k)!}\sqrt{m!}\sqrt{(m+k)!}} \\
& \times e^{-\frac{1}{2}(|\alpha_p|^2+|\bar{\alpha}_p|^2+|\alpha_S|^2+|\bar{\alpha}_S|^2)}
  A_{n,m,k}(t),  \nonumber 
\end{align}
and $\bar{\alpha}_p$ and $\bar{\alpha}_S$ are the measured complex amplitudes for the pump and Stokes modes. 
For a given experimental configuration (i.e., input laser amplitudes and measured final states), examination of the phonon Fock state probabilities as a function of time identify how certain states can be prepared (e.g., see Fig. \ref{fig:wigHomodyne}(a)). Such conditional measurements can collapse the phonon into a highly nonclassical state as shown in Fig. \ref{fig:wigHomodyne}(b)-(e). For example, in Fig. \ref{fig:wigHomodyne}(b) the system occupies a nearly perfect superposition $|\psi_{ph}\rangle  \approx (|2\rangle + |3\rangle \exp\{i\phi\})/\sqrt{2}$. Accounting for decoherence caused by optical losses, a master equation simulation \cite{johansson2012qutip} shows that this state can be achieved with 72\% fidelity when the optical decay rates $\gamma = 0.25 g$ (see Supplementary Information \ref{MEsim}). With the large optomechanical coupling rates achievable in optomechanical crystals \cite{Chan2009} ($\sim$ 9 MHz), it is conceivable that future devices reach regimes where $\gamma \ll g$. However, even with larger optical losses, interesting states can still be accessed. By adjusting the initial optical state, entanglement generation can be accelerated so that targeted quantum states can be prepared at times $t < 1/\gamma$ (example below). Beyond this example, a tremendous range of nonclassical states become accessible by varying the amplitude and phases of initial and final photonic states as well as the measurement time. 

We can also use Eq. \eqref{WF} to identify opportunities for quantum state synthesis using conditional photon number resolving measurements.
For example, when $|\psi(0)\rangle = |\alpha_p, \alpha_S,0\rangle$ and when a single photon is measured in the pump and Stokes modes at time $t$, the phonon wavefunction collapses into the superposition of ground and excited states, expressed in unnormalized form as
\begin{align}
\label{phNRM}
| \psi_{ph}[1_p,1_s,t) \rangle \!=\!  \mathcal{N}\big[\alpha_S \cos(\sqrt{2} gt)  |0 \rangle \!-\! \frac{i\alpha_p}{\sqrt{6}}  \sin(\sqrt{6} gt)  |1 \rangle\big]
\end{align}
where $\mathcal{N} = \alpha_p\mathcal{P}_0\mathcal{S}_0$. Although nondeterministic, manipulation of the initial state affords some control over the phonon wavefunction, e.g., by setting the amplitude $\alpha_S$ to zero the mechanical mode is guaranteed to be in a single phonon state. 

\begin{figure}[h]
    \centering
    \includegraphics[width = 0.45\textwidth]{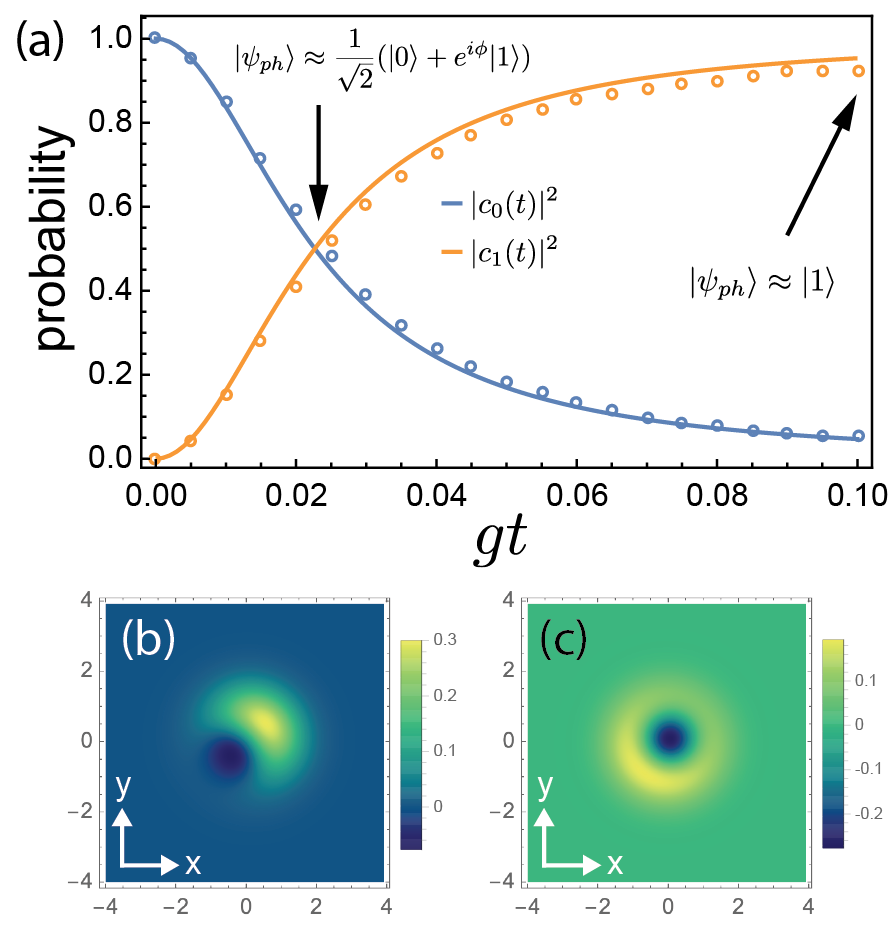}
    \vspace{-0.1in}
    \caption{Phonon Fock state probabilities (a) and Wigner functions (b)-(c) after single photon measurement in the Stokes mode and homodyne measurement in the pump mode. Fock state probabilities computed using Eq. \eqref{WF} (lines) and a master equation solver including optical losses of decay rate $\gamma = 50 g$ (open circles). Plotted Wigner functions include the effects of optical losses. The inital state is $|\alpha_p,\alpha_S,0\rangle$ with $\alpha_p = 4.41$ and $\alpha_S = 0.1$, and the measured complex amplitude of the pump is $\bar{\alpha}_p = 3.7$.}
    \label{fig:entAcc}
    \vspace{-0.1in}
\end{figure}

Using single photon detection, we can illustrate how quantum state synthesis can be achieved with optical losses. Figure \ref{fig:entAcc} shows the phonon Fock state probabilities, computed using Eq. \eqref{WF} (solid lines) and a master equation solver (open circles), along with the phonon Wigner functions for the simulation with optical losses at selected times when a single Stokes photon is measured and the pump is measured using homodyne detection. This figure shows that even with optical decay rates of $50g$, a coherent superposition between ground and excited states and a single phonon state can be prepared with respective fidelities of 97\% and 92\%. 

\begin{figure}[ht]
\includegraphics[width=8.5cm]{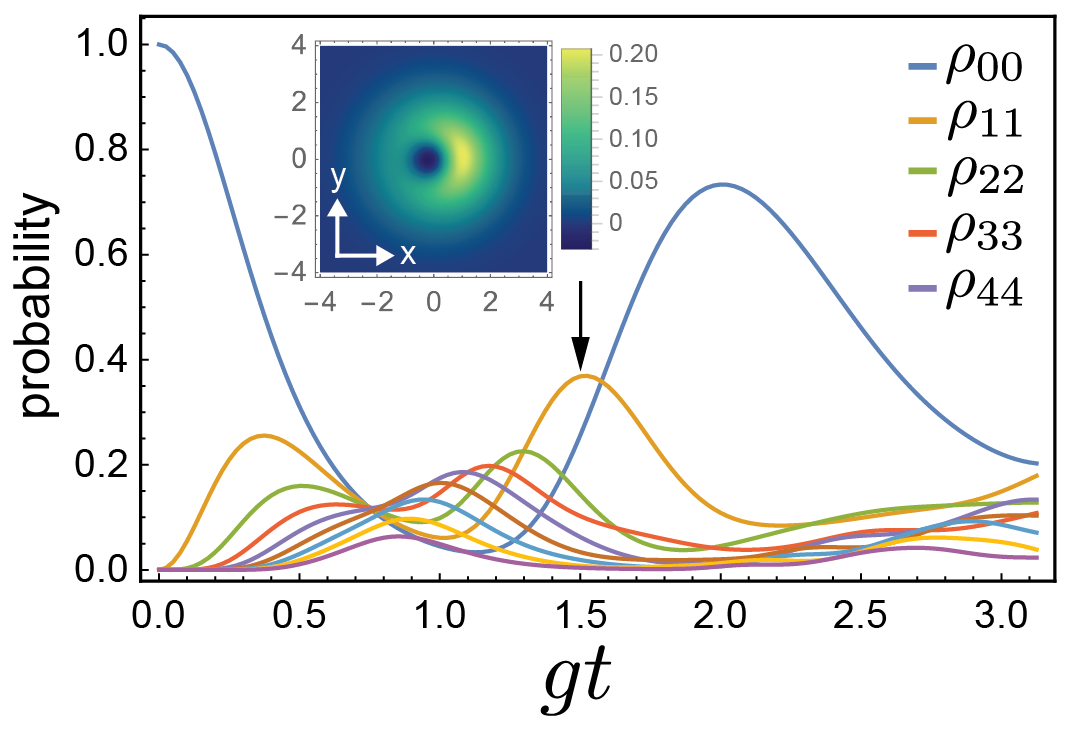}
\begin{center}
\vspace{-0.25 in}
\caption{Phonon occupation number probabilities when optical fields are traced out (i.e., not measured). The initial state is $|\alpha_p, \alpha_S, 0 \rangle$ with $\alpha_p = 2.4$ and $\alpha_S = 0.25$. Inset: phonon Wigner function at $gt=1.5$.}
\label{WigF2}
\vspace{-0.35 in}
\end{center}
\end{figure}

Even in the absence of conditional measurements, the mechanical oscillator can evolve into nonclassical states, leading to a form of deterministic state synthesis. In this case, the phonon mode is described by a reduced density matrix $\hat{\rho}_{red}(t) = \sum_{k,k'=0}^\infty \rho_{kk'}(t) |k\rangle \langle k'|$ with matrix element given by 
\begin{align}
\rho_{kk'}(t)=\sum_{n=0}^\infty \sum_{m=\kappa}^\infty & \mathcal{P}_{n+k} \mathcal{P}^*_{n+k'} \mathcal{S}_{m-k} \mathcal{S}^*_{m-k'} 
 \\
& \times  A_{n+k,m-k,k}(t) A^*_{n+k',m-k',k'}(t) \nonumber
\end{align}
where $\kappa ={\rm max}(k,k')$. For a specific initial state, Fig. \ref{WigF2} (a) illustrates how the Fock state probabilities for the phonon evolve in time, showing how a variety of macroscopic superposition states can be prepared Fig. \ref{WigF2}. We emphasize that this behavior is a consequence of the quantum nonlinearity. Importantly, without the effects backreaction on the pump and the Stokes modes (e.g., pump depletion) provided by the nonlinear dynamics produced by Eq. \eqref{H}, the phonon would evolve into a thermal or coherent state \cite{gerry2005introductory}, and negativity of the Wigner function would not be possible. 

{\it Quantum state readout}: In addition to providing a means to prepare quantum states of mechanical motion, $H_{int}$ also enables state manipulation, transfer and readout. When the Stokes mode is a large amplitude coherent state (i.e., $|\alpha_S| \gg 1$), the $\exp\{-i H_{int} t/\hbar\}$ is well-approximated by the beam splitter transformation, transferring and entangling quantum states between the phononic and photonic domains \cite{aspelmeyer2014cavity}. Consider a phonon wavefunction given by a superposition of ground and excited states $|\psi_{ph}\rangle = c_0 |0\rangle + c_1 |1\rangle$. Equation \eqref{phNRM} shows how this state can be  prepared by conditional photon number resolving measurements. 
For the pump mode in the vacuum state and Stokes mode in a coherent state, the system wavefunction at time $t$ is 
\begin{align}
\label{swap}
| \psi(t) \rangle = \sum_{m=0}^\infty \mathcal{S}_m \bigg[& c_0 |0,m,0\rangle + c_1\cos(\sqrt{m} gt) |0,m,1\rangle\nonumber 
\\& -i c_1\sin(\sqrt{m} gt) |1,m-1,0\rangle \bigg].
\end{align}
Under these assumptions, $\mathcal{S}_m$ is peaked at $m \sim |\alpha_S|^2$ with a standard deviation $\Delta m = |\alpha_S|$, allowing the arguments of the cosine and sine terms to be approximated by $\sqrt{m} gt \approx |\alpha_S|gt$ when $gt \ll 1$. Assuming the phase of $\alpha_S$ is $\pi/2$, these approximations give
\begin{align}
\label{swap}
| \psi(t) \rangle \approx  c_0 |0,\alpha_S,0\rangle +c_1& \big[\cos(|\alpha_S| gt) |0,\alpha_S,1\rangle\nonumber 
\\ & + \sin(|\alpha_S| gt) |1,\alpha_S,0\rangle\big],
\end{align}
providing a ``state-swap" when $ t = t_\pi  \equiv \pi/(2|\alpha_S|g)$. For this generalized tripartite optomechanical ``$\pi$-pulse'', $|\psi(t_\pi) \rangle \approx  (c_0 |0\rangle + c_1 |1\rangle) \otimes|\alpha_S,0\rangle$ showing how a quantum state in the mechanical domain can be transferred to the optical domain where quantum state tomography can be performed. This result also shows how an optomechanical $\pi/2$-pulse ($|\alpha|gt_{\pi/2} = \pi/4$) can be used to prepare an entangled superposition of the pump and phonon. For $c_0 =0$ and $c_1 = 1$, a phonon-photon Bell state is produced
\begin{align}
\label{swap}
| \psi(t_{\pi/2}) \rangle \!\approx\!  \frac{1}{\sqrt{2}}& \big[|0,\alpha_S,1\rangle\!-\! i e^{i\varphi} |1,\alpha_S,0\rangle\big]
\end{align}
where $\varphi$ is the phase of the Stokes mode. By sequencing these pulses, multimode optomechanical analogs of Ramsey interference and spin echo can be performed. While these results, focus on the $|\alpha_S| \gg 1$ limit, Eq. \eqref{swap} can be used to describe general ``state-swap" dynamics. 
 
{\it Conclusion}: 
In this letter we have shown how three-mode optomechanical coupling can be used to prepare and measure quantum states of a mechanical oscillator. A formal solution to the Schrodinger equation for this system yields highly entangled photon-phonon states, even with classically prepared initial states. Levering this entanglement, conditional measurements on the optical fields can generate a wide variety of quantum states of the mechanical oscillator. For quantum state tomography, this tripartite coupling permits the optomechanical equivalent of $\pi/2$- and $\pi$-pulses that can be used to entangle optical and mechanical states as well as transfer quantum information between the electromagnetic and mechanical domains. With the long lifetimes at cryogenic temperatures, natural interface with telecommunications and radio frequency wavelengths, and large effective masses of phonon modes, our results may enable new ways to manipulate quantum information stored in mechanical oscillators as well as tests of the foundations of quantum physics. 

This work was supported by NSF Award 2145724 and !`MIRA! ER Funds. The authors thank In$\rm{\grave{e}}$s Monta$\rm{\tilde{n}}$o,  Daniel Reiche, and Shruti Puri for stimulating discussions. 

\appendix

\section{Formal equivalence between the multimode optomechanical Hamiltonian and the Jaynes-Cummings model}
\label{SI:FormalEquiv}
Leveraging the conservation of the number of pump photons and phonons and the Schwinger oscillator model of angular momentum \cite{Schwinger1952}, a formal analogy between the multimode optomechanical Hamiltonian and the Jaynes-Cummings model \cite{jaynes1963comparison} can be derived. To reveal this relationship, we define the following pseudospin operators
\begin{align}
& \hat{N} =  (\hat{n}_p + \hat{n}_b)/2 \\
& S_z =  (\hat{n}_p - \hat{n}_b)/2 \\
& S_- =  a_p b^\dag \\
& S_+ =  a_p^\dag b
\end{align}
where $\hat{n}_p$ and $\hat{n}_b$ are the number operators for the pump and phonon modes respectively. 
The commutation relations of $a_p$ and $b$ show that the pseudospin operators obey the relations
\begin{align}
& [\hat{N},S_z]  =  0 \\
& [S_z,S_\pm] =  \pm S_\pm \\
& [S_+,S_-] = 2S_z. 
\end{align}
Furthermore, one can show that $N$, the eigenvalue for $\hat{N}$, plays the role for the total spin, that these operators have all of the expected effects on Fock states $|n_p,n_b\rangle$ where the eigenvalues of $S_z$, $m \equiv (n_p-n_b)/2$, correspond with azimuthal component of the pseudospin along the ``z-axis". For example, using $N$ and $m$, a generic Fock state $|n_p,n_b\rangle$ can be formally expressed as $| N,m\rangle $ ($n_p = N+m$ and $n_b = N-m$) where the operators above have act in a manner analogous to a spin system:
\begin{align}
S_z | N,m \rangle   & =  m| N,m \rangle \\
S_\pm | N,m\rangle & =  \sqrt{N\pm m+1}\sqrt{N \mp m} | N,m\pm1\rangle \\
{\bf S}^2 | N,m\rangle & = (S_z^2-S_z+S_+S_-) | N,m\rangle \\
& =  N\left(N+1\right)| N,m, j\rangle. 
\end{align}

Using these operators and the phase matching conditions one can show that 
\begin{align}
 H   & \to  \hbar (\omega_p+\Omega) N + \hbar \omega_S S_z + \hbar \omega_S a^\dag_S a_S  +  \hbar g( S_- a^\dag_S  + S_+ a_S )
\end{align}
demonstrating the equivalence between the multimode Hamiltonian and the JC model when $N$ is fixed. In general, initial states that can be prepared in the lab will contain a superposition of different $N$-states, such as coherent states. Therefore, the multimode optomechanical dynamics of realistic systems is complex, involving several pseudospins, with distinct spin $N$, interacting with a bosonic mode. 

\section{Diagonalization of ${\bf M}_{nm}$}
\label{DiagonalM}
The formal solution to Eq. \eqref{probAmps} can be obtained by diagonalizing ${\bf M}_{nm}$. By solving the eigenvalue equation
${\bf M}_{nm}\cdot \vec{E}_{nm,k} = {\Omega}_{nm,k} \vec{E}_{nm,k}$, where $\vec{E}^*_{nm,k}\cdot \vec{E}_{nm,k'} = \delta_{k,k'}$, the eigenvalues can be obtained as well as the unitary matrix ${\bf V}_{nm}$.  Comprising a matrix of the normalized eigenvectors of ${\bf M}_{nm}$, the definition of ${\bf V}_{nm}$
\begin{eqnarray}
\label{}
{\bf V}_{nm}
= 
\begin{pmatrix}
\vec{E}_{nm,n},
&
\vec{E}_{nm,n-1},
&
\hdots
&
\vec{E}_{nm,0}
\end{pmatrix}.
\end{eqnarray}
For example, for $n_p+n_b =1$, the normalized eigenvectors are
\begin{align}
\vec{E}_{1m,1} = \frac{1}{\sqrt{2}}
\begin{pmatrix}
       1   \\
        1 
\end{pmatrix}, \quad
\vec{E}_{1m,0} = \frac{1}{\sqrt{2}}
\begin{pmatrix}
       -1   \\
        1 
\end{pmatrix}
\end{align}
giving ${\bf V}_{1m}$ 
\begin{align}
{\bf V}_{1m} = \frac{1}{\sqrt{2}}
\begin{pmatrix}
       1 & -1  \\
        1 & 1
        \end{pmatrix}.
\end{align}
Using ${\bf V}_{1m}$ to diagonalize ${\bf M}_{1m}$ we obtain the diagonal matrix of eigenvalues ${\bf \Omega}_{1m}$  
\begin{align}
{\bf V}_{1m}^\dag\cdot{\bf M}_{1m}\cdot {\bf V}_{1m} \equiv {\bf \Omega}_{1m}= g\sqrt{m+1}
\begin{pmatrix}
       1 & 0  \\
        0 & -1
        \end{pmatrix}.
\end{align}

\section{Probability amplitude time-dynamics}
\label{sec:probabilityAmplitudes}
Here we list the first few terms $A_{n,m,k}(t)$. By solving Eq. \eqref{probAmps}, assuming the phonon is initially in the ground state, we find

\begin{align}
& A_{0,m,0}(t) = 1 \\
&A_{1,m,0}(t) = \cos (\sqrt{m+1}gt) \\
&A_{1,m,1}(t) = -i  \sin (\sqrt{m+1}gt)1 \\
&A_{2,m,0}(t) = \frac{2+m+(1+m)\cos(\sqrt{4m +6} gt)}{3+2m} \\
&A_{2,m,1}(t) =- i \sqrt{\frac{1+m}{3+2m}}\sin(\sqrt{4 m+6} gt) \\
&A_{2,m,2}(t) = -2\frac{\sqrt{(1+m)(2+m)}}{3+2m}\sin^2(\sqrt{m+3/2} gt) 
\end{align}

\section{General Expression for the Wigner function}
\label{SI:WignerFunction}
The phonon Wigner function is given by 
\begin{align}
\label{eq:wigner}
W(\alpha) = \frac{1}{2\pi^2} \int d^2 \xi \ e^{\alpha \xi^*-\alpha^* \xi} {\rm Tr}[\hat{\rho} D(\xi)]
\end{align}
where $\alpha \equiv (x+iy)/\sqrt{2}$, $x$ and $y$ are dimensionless position and momentum quadratures, and $D(\xi)$ is the phonon displacement operator \cite{gerry2005introductory}. Here, $\hat{\rho}$ is either reduced phonon density matrix, or the density matrix obtained after conditional measurement of the optical fields. 

For a general state expressed in the Fock basis (i.e., $\rho_{kk'}$), the Wigner function is given by 
\begin{align}
W(\alpha) = \frac{1}{\pi} \sum_{k,k'=0}^\infty \frac{(-1)^k 2^{k'-k}}{\sqrt{k!}\sqrt{k'!}} (\partial_\alpha\!-\! 2\alpha^*)^k \alpha^{k'} e^{-2 \alpha \alpha^*} 
\rho_{kk'}
\end{align}
where $\rho_{k k'} = c_k c_{k'}^*$ for a pure state and $c_k$ is the phonon probability amplitude for the $k$th Fock state.

For the case where the optical fields are conditionally measured, the general expression for the phonon Wigner function can be computed by making the following replacement in Eq. \eqref{eq:wigner} is given by 
\begin{align}
{\rm Tr}[\hat{\rho} D(\xi)] \to \langle \psi_{ph}[\varphi_p,\varphi_S,t)| D(\xi) | \psi_{ph}[\varphi_p,\varphi_S,t) \rangle.
\end{align}

\section{Effects of optical losses from master equation simulations}
\label{MEsim}
For the assessment of decoherence caused by optical losses, we simulate the quantum dynamics of this tripartite optomechanical system \cite{johansson2012qutip}. We solve the Linblad form of the the master equation, capturing the decay of both optical modes at a rate $\gamma$ given by
\begin{equation}
\label{LME}
    \dot{\rho} = -\frac{i}{\hbar} [H,\rho] +\frac{\gamma}{2} \sum_{j=p,S}[2a_j\rho a^\dag_j - a^\dag_ja_j \rho- \rho a^\dag_ja_j]. 
\end{equation}
Given the high frequency of the photon modes, Eq. \eqref{LME} assumes a bath temperature of zero.


\begin{thebibliography}{43}%
\makeatletter
\providecommand \@ifxundefined [1]{%
 \@ifx{#1\undefined}
}%
\providecommand \@ifnum [1]{%
 \ifnum #1\expandafter \@firstoftwo
 \else \expandafter \@secondoftwo
 \fi
}%
\providecommand \@ifx [1]{%
 \ifx #1\expandafter \@firstoftwo
 \else \expandafter \@secondoftwo
 \fi
}%
\providecommand \natexlab [1]{#1}%
\providecommand \enquote  [1]{``#1''}%
\providecommand \bibnamefont  [1]{#1}%
\providecommand \bibfnamefont [1]{#1}%
\providecommand \citenamefont [1]{#1}%
\providecommand \href@noop [0]{\@secondoftwo}%
\providecommand \href [0]{\begingroup \@sanitize@url \@href}%
\providecommand \@href[1]{\@@startlink{#1}\@@href}%
\providecommand \@@href[1]{\endgroup#1\@@endlink}%
\providecommand \@sanitize@url [0]{\catcode `\\12\catcode `\$12\catcode
  `\&12\catcode `\#12\catcode `\^12\catcode `\_12\catcode `\%12\relax}%
\providecommand \@@startlink[1]{}%
\providecommand \@@endlink[0]{}%
\providecommand \url  [0]{\begingroup\@sanitize@url \@url }%
\providecommand \@url [1]{\endgroup\@href {#1}{\urlprefix }}%
\providecommand \urlprefix  [0]{URL }%
\providecommand \Eprint [0]{\href }%
\providecommand \doibase [0]{http://dx.doi.org/}%
\providecommand \selectlanguage [0]{\@gobble}%
\providecommand \bibinfo  [0]{\@secondoftwo}%
\providecommand \bibfield  [0]{\@secondoftwo}%
\providecommand \translation [1]{[#1]}%
\providecommand \BibitemOpen [0]{}%
\providecommand \bibitemStop [0]{}%
\providecommand \bibitemNoStop [0]{.\EOS\space}%
\providecommand \EOS [0]{\spacefactor3000\relax}%
\providecommand \BibitemShut  [1]{\csname bibitem#1\endcsname}%
\let\auto@bib@innerbib\@empty


\bibitem [{\citenamefont {Marshall}\ \emph {et~al.}(2003)\citenamefont
  {Marshall}, \citenamefont {Simon}, \citenamefont {Penrose},\ and\
  \citenamefont {Bouwmeester}}]{Marshall2003}%
  \BibitemOpen
  \bibfield  {author} {\bibinfo {author} {\bibfnamefont {W.}~\bibnamefont
  {Marshall}}, \bibinfo {author} {\bibfnamefont {C.}~\bibnamefont {Simon}},
  \bibinfo {author} {\bibfnamefont {R.}~\bibnamefont {Penrose}}, \ and\
  \bibinfo {author} {\bibfnamefont {D.}~\bibnamefont {Bouwmeester}},\
  }\href@noop {} {\bibfield  {journal} {\bibinfo  {journal} {Physical Review
  Letters}\ }\textbf {\bibinfo {volume} {91}},\ \bibinfo {pages} {130401}
  (\bibinfo {year} {2003})}\BibitemShut {NoStop}%
\bibitem [{\citenamefont {Goryachev}\ \emph {et~al.}(2012)\citenamefont
  {Goryachev}, \citenamefont {Creedon}, \citenamefont {Ivanov}, \citenamefont
  {Galliou}, \citenamefont {Bourquin},\ and\ \citenamefont
  {Tobar}}]{Goryachev2012}%
  \BibitemOpen
  \bibfield  {author} {\bibinfo {author} {\bibfnamefont {M.}~\bibnamefont
  {Goryachev}}, \bibinfo {author} {\bibfnamefont {D.~L.}\ \bibnamefont
  {Creedon}}, \bibinfo {author} {\bibfnamefont {E.~N.}\ \bibnamefont {Ivanov}},
  \bibinfo {author} {\bibfnamefont {S.}~\bibnamefont {Galliou}}, \bibinfo
  {author} {\bibfnamefont {R.}~\bibnamefont {Bourquin}}, \ and\ \bibinfo
  {author} {\bibfnamefont {M.~E.}\ \bibnamefont {Tobar}},\ }\href@noop {}
  {\bibfield  {journal} {\bibinfo  {journal} {Applied Physics Letters}\
  }\textbf {\bibinfo {volume} {100}},\ \bibinfo {pages} {243504} (\bibinfo
  {year} {2012})}\BibitemShut {NoStop}%
\bibitem [{\citenamefont {Goryachev}\ \emph {et~al.}(2013)\citenamefont
  {Goryachev}, \citenamefont {Creedon}, \citenamefont {Galliou},\ and\
  \citenamefont {Tobar}}]{Goryachev2013}%
  \BibitemOpen
  \bibfield  {author} {\bibinfo {author} {\bibfnamefont {M.}~\bibnamefont
  {Goryachev}}, \bibinfo {author} {\bibfnamefont {D.~L.}\ \bibnamefont
  {Creedon}}, \bibinfo {author} {\bibfnamefont {S.}~\bibnamefont {Galliou}}, \
  and\ \bibinfo {author} {\bibfnamefont {M.~E.}\ \bibnamefont {Tobar}},\
  }\href@noop {} {\bibfield  {journal} {\bibinfo  {journal} {Physical Review
  Letters}\ }\textbf {\bibinfo {volume} {111}},\ \bibinfo {pages} {085502}
  (\bibinfo {year} {2013})}\BibitemShut {NoStop}%
\bibitem [{\citenamefont {Galliou}\ \emph {et~al.}(2013)\citenamefont
  {Galliou}, \citenamefont {Goryachev}, \citenamefont {Bourquin}, \citenamefont
  {Abb{\'{e}}}, \citenamefont {Aubry},\ and\ \citenamefont
  {Tobar}}]{Galliou2013a}%
  \BibitemOpen
  \bibfield  {author} {\bibinfo {author} {\bibfnamefont {S.}~\bibnamefont
  {Galliou}}, \bibinfo {author} {\bibfnamefont {M.}~\bibnamefont {Goryachev}},
  \bibinfo {author} {\bibfnamefont {R.}~\bibnamefont {Bourquin}}, \bibinfo
  {author} {\bibfnamefont {P.}~\bibnamefont {Abb{\'{e}}}}, \bibinfo {author}
  {\bibfnamefont {J.~P.}\ \bibnamefont {Aubry}}, \ and\ \bibinfo {author}
  {\bibfnamefont {M.~E.}\ \bibnamefont {Tobar}},\ }\href@noop {} {\bibfield
  {journal} {\bibinfo  {journal} {Scientific reports}\ }\textbf {\bibinfo
  {volume} {3}},\ \bibinfo {pages} {2132} (\bibinfo {year} {2013})}\BibitemShut
  {NoStop}%
\bibitem [{\citenamefont {Lo}\ \emph {et~al.}(2016)\citenamefont {Lo},
  \citenamefont {Haslinger}, \citenamefont {Mizrachi}, \citenamefont
  {Anderegg}, \citenamefont {M{\"u}ller}, \citenamefont {Hohensee},
  \citenamefont {Goryachev},\ and\ \citenamefont {Tobar}}]{Lo2016}%
  \BibitemOpen
  \bibfield  {author} {\bibinfo {author} {\bibfnamefont {A.}~\bibnamefont
  {Lo}}, \bibinfo {author} {\bibfnamefont {P.}~\bibnamefont {Haslinger}},
  \bibinfo {author} {\bibfnamefont {E.}~\bibnamefont {Mizrachi}}, \bibinfo
  {author} {\bibfnamefont {L.}~\bibnamefont {Anderegg}}, \bibinfo {author}
  {\bibfnamefont {H.}~\bibnamefont {M{\"u}ller}}, \bibinfo {author}
  {\bibfnamefont {M.}~\bibnamefont {Hohensee}}, \bibinfo {author}
  {\bibfnamefont {M.}~\bibnamefont {Goryachev}}, \ and\ \bibinfo {author}
  {\bibfnamefont {M.~E.}\ \bibnamefont {Tobar}},\ }\href@noop {} {\bibfield
  {journal} {\bibinfo  {journal} {Physical Review X}\ }\textbf {\bibinfo
  {volume} {6}},\ \bibinfo {pages} {011018} (\bibinfo {year}
  {2016})}\BibitemShut {NoStop}%
\bibitem [{\citenamefont {Tobar}(2017)}]{Tobar2017}%
  \BibitemOpen
  \bibfield  {author} {\bibinfo {author} {\bibfnamefont {M.~E.}\ \bibnamefont
  {Tobar}},\ }\href@noop {} {\bibfield  {journal} {\bibinfo  {journal} {New
  Journal of Physics}\ }\textbf {\bibinfo {volume} {19}},\ \bibinfo {pages}
  {091001} (\bibinfo {year} {2017})}\BibitemShut {NoStop}%
\bibitem [{\citenamefont {Renninger}\ \emph {et~al.}(2018)\citenamefont
  {Renninger}, \citenamefont {Kharel}, \citenamefont {Behunin},\ and\
  \citenamefont {Rakich}}]{Renninger2018}%
  \BibitemOpen
  \bibfield  {author} {\bibinfo {author} {\bibfnamefont {W.}~\bibnamefont
  {Renninger}}, \bibinfo {author} {\bibfnamefont {P.}~\bibnamefont {Kharel}},
  \bibinfo {author} {\bibfnamefont {R.}~\bibnamefont {Behunin}}, \ and\
  \bibinfo {author} {\bibfnamefont {P.}~\bibnamefont {Rakich}},\ }\href
  {\doibase 10.1038/s41567-018-0090-3} {\bibfield  {journal} {\bibinfo
  {journal} {Nature Physics}\ } (\bibinfo {year} {2018}),\
  10.1038/s41567-018-0090-3}\BibitemShut {NoStop}%
\bibitem [{\citenamefont {Kharel}\ \emph {et~al.}(2018)\citenamefont {Kharel},
  \citenamefont {Chu}, \citenamefont {Power}, \citenamefont {Renninger},
  \citenamefont {Schoelkopf},\ and\ \citenamefont {Rakich}}]{Kharel2018}%
  \BibitemOpen
  \bibfield  {author} {\bibinfo {author} {\bibfnamefont {P.}~\bibnamefont
  {Kharel}}, \bibinfo {author} {\bibfnamefont {Y.}~\bibnamefont {Chu}},
  \bibinfo {author} {\bibfnamefont {M.}~\bibnamefont {Power}}, \bibinfo
  {author} {\bibfnamefont {W.~H.}\ \bibnamefont {Renninger}}, \bibinfo {author}
  {\bibfnamefont {R.~J.}\ \bibnamefont {Schoelkopf}}, \ and\ \bibinfo {author}
  {\bibfnamefont {P.~T.}\ \bibnamefont {Rakich}},\ }\href@noop {} {\bibfield
  {journal} {\bibinfo  {journal} {APL Photonics}\ }\textbf {\bibinfo {volume}
  {3}},\ \bibinfo {pages} {066101} (\bibinfo {year} {2018})}\BibitemShut
  {NoStop}%
\bibitem [{\citenamefont {Eichenfield}\ \emph
  {et~al.}(2009{\natexlab{a}})\citenamefont {Eichenfield}, \citenamefont
  {Camacho}, \citenamefont {Chan}, \citenamefont {Vahala},\ and\ \citenamefont
  {Painter}}]{Eichenfield2009}%
  \BibitemOpen
  \bibfield  {author} {\bibinfo {author} {\bibfnamefont {M.}~\bibnamefont
  {Eichenfield}}, \bibinfo {author} {\bibfnamefont {R.}~\bibnamefont
  {Camacho}}, \bibinfo {author} {\bibfnamefont {J.}~\bibnamefont {Chan}},
  \bibinfo {author} {\bibfnamefont {K.~J.}\ \bibnamefont {Vahala}}, \ and\
  \bibinfo {author} {\bibfnamefont {O.}~\bibnamefont {Painter}},\ }\href@noop
  {} {\bibfield  {journal} {\bibinfo  {journal} {Nature}\ }\textbf {\bibinfo
  {volume} {459}},\ \bibinfo {pages} {550} (\bibinfo {year}
  {2009}{\natexlab{a}})}\BibitemShut {NoStop}%
\bibitem [{\citenamefont {Eichenfield}\ \emph
  {et~al.}(2009{\natexlab{b}})\citenamefont {Eichenfield}, \citenamefont
  {Chan}, \citenamefont {Camacho}, \citenamefont {Vahala},\ and\ \citenamefont
  {Painter}}]{Eichenfield2009a}%
  \BibitemOpen
  \bibfield  {author} {\bibinfo {author} {\bibfnamefont {M.}~\bibnamefont
  {Eichenfield}}, \bibinfo {author} {\bibfnamefont {J.}~\bibnamefont {Chan}},
  \bibinfo {author} {\bibfnamefont {R.~M.}\ \bibnamefont {Camacho}}, \bibinfo
  {author} {\bibfnamefont {K.~J.}\ \bibnamefont {Vahala}}, \ and\ \bibinfo
  {author} {\bibfnamefont {O.}~\bibnamefont {Painter}},\ }\href@noop {}
  {\bibfield  {journal} {\bibinfo  {journal} {Nature}\ }\textbf {\bibinfo
  {volume} {462}},\ \bibinfo {pages} {78} (\bibinfo {year}
  {2009}{\natexlab{b}})}\BibitemShut {NoStop}%
\bibitem [{\citenamefont {Weaver}\ \emph {et~al.}(2017)\citenamefont {Weaver},
  \citenamefont {Buters}, \citenamefont {Luna}, \citenamefont {Eerkens},
  \citenamefont {Heeck}, \citenamefont {de~Man},\ and\ \citenamefont
  {Bouwmeester}}]{weaver2017}%
  \BibitemOpen
  \bibfield  {author} {\bibinfo {author} {\bibfnamefont {M.~J.}\ \bibnamefont
  {Weaver}}, \bibinfo {author} {\bibfnamefont {F.}~\bibnamefont {Buters}},
  \bibinfo {author} {\bibfnamefont {F.}~\bibnamefont {Luna}}, \bibinfo {author}
  {\bibfnamefont {H.}~\bibnamefont {Eerkens}}, \bibinfo {author} {\bibfnamefont
  {K.}~\bibnamefont {Heeck}}, \bibinfo {author} {\bibfnamefont
  {S.}~\bibnamefont {de~Man}}, \ and\ \bibinfo {author} {\bibfnamefont
  {D.}~\bibnamefont {Bouwmeester}},\ }\href@noop {} {\bibfield  {journal}
  {\bibinfo  {journal} {Nature communications}\ }\textbf {\bibinfo {volume}
  {8}},\ \bibinfo {pages} {1} (\bibinfo {year} {2017})}\BibitemShut {NoStop}%
\bibitem [{\citenamefont {Campbell}\ \emph {et~al.}(2021)\citenamefont
  {Campbell}, \citenamefont {McAllister}, \citenamefont {Goryachev},
  \citenamefont {Ivanov},\ and\ \citenamefont {Tobar}}]{campbell2021}%
  \BibitemOpen
  \bibfield  {author} {\bibinfo {author} {\bibfnamefont {W.~M.}\ \bibnamefont
  {Campbell}}, \bibinfo {author} {\bibfnamefont {B.~T.}\ \bibnamefont
  {McAllister}}, \bibinfo {author} {\bibfnamefont {M.}~\bibnamefont
  {Goryachev}}, \bibinfo {author} {\bibfnamefont {E.~N.}\ \bibnamefont
  {Ivanov}}, \ and\ \bibinfo {author} {\bibfnamefont {M.~E.}\ \bibnamefont
  {Tobar}},\ }\href@noop {} {\bibfield  {journal} {\bibinfo  {journal}
  {Physical Review Letters}\ }\textbf {\bibinfo {volume} {126}},\ \bibinfo
  {pages} {071301} (\bibinfo {year} {2021})}\BibitemShut {NoStop}%
\bibitem [{\citenamefont {MacCabe}\ \emph {et~al.}(2020)\citenamefont
  {MacCabe}, \citenamefont {Ren}, \citenamefont {Luo}, \citenamefont {Cohen},
  \citenamefont {Zhou}, \citenamefont {Sipahigil}, \citenamefont
  {Mirhosseini},\ and\ \citenamefont {Painter}}]{maccabe2020nano}%
  \BibitemOpen
  \bibfield  {author} {\bibinfo {author} {\bibfnamefont {G.~S.}\ \bibnamefont
  {MacCabe}}, \bibinfo {author} {\bibfnamefont {H.}~\bibnamefont {Ren}},
  \bibinfo {author} {\bibfnamefont {J.}~\bibnamefont {Luo}}, \bibinfo {author}
  {\bibfnamefont {J.~D.}\ \bibnamefont {Cohen}}, \bibinfo {author}
  {\bibfnamefont {H.}~\bibnamefont {Zhou}}, \bibinfo {author} {\bibfnamefont
  {A.}~\bibnamefont {Sipahigil}}, \bibinfo {author} {\bibfnamefont
  {M.}~\bibnamefont {Mirhosseini}}, \ and\ \bibinfo {author} {\bibfnamefont
  {O.}~\bibnamefont {Painter}},\ }\href@noop {} {\bibfield  {journal} {\bibinfo
   {journal} {Science}\ }\textbf {\bibinfo {volume} {370}},\ \bibinfo {pages}
  {840} (\bibinfo {year} {2020})}\BibitemShut {NoStop}%
\bibitem [{\citenamefont {Pechal}\ \emph {et~al.}(2018)\citenamefont {Pechal},
  \citenamefont {Arrangoiz-Arriola},\ and\ \citenamefont
  {Safavi-Naeini}}]{pechal2019}%
  \BibitemOpen
  \bibfield  {author} {\bibinfo {author} {\bibfnamefont {M.}~\bibnamefont
  {Pechal}}, \bibinfo {author} {\bibfnamefont {P.}~\bibnamefont
  {Arrangoiz-Arriola}}, \ and\ \bibinfo {author} {\bibfnamefont {A.~H.}\
  \bibnamefont {Safavi-Naeini}},\ }\href@noop {} {\bibfield  {journal}
  {\bibinfo  {journal} {Quantum Science and Technology}\ }\textbf {\bibinfo
  {volume} {4}},\ \bibinfo {pages} {015006} (\bibinfo {year}
  {2018})}\BibitemShut {NoStop}%
\bibitem [{\citenamefont {Chu}\ \emph {et~al.}(2017)\citenamefont {Chu},
  \citenamefont {Kharel}, \citenamefont {Renninger}, \citenamefont {Burkhart},
  \citenamefont {Frunzio}, \citenamefont {Rakich},\ and\ \citenamefont
  {Schoelkopf}}]{Chu2017}%
  \BibitemOpen
  \bibfield  {author} {\bibinfo {author} {\bibfnamefont {Y.}~\bibnamefont
  {Chu}}, \bibinfo {author} {\bibfnamefont {P.}~\bibnamefont {Kharel}},
  \bibinfo {author} {\bibfnamefont {W.~H.}\ \bibnamefont {Renninger}}, \bibinfo
  {author} {\bibfnamefont {L.~D.}\ \bibnamefont {Burkhart}}, \bibinfo {author}
  {\bibfnamefont {L.}~\bibnamefont {Frunzio}}, \bibinfo {author} {\bibfnamefont
  {P.~T.}\ \bibnamefont {Rakich}}, \ and\ \bibinfo {author} {\bibfnamefont
  {R.~J.}\ \bibnamefont {Schoelkopf}},\ }\href@noop {} {\bibfield  {journal}
  {\bibinfo  {journal} {Science}\ }\textbf {\bibinfo {volume} {358}},\ \bibinfo
  {pages} {199} (\bibinfo {year} {2017})}\BibitemShut {NoStop}%
\bibitem [{\citenamefont {Chu}\ \emph {et~al.}(2018)\citenamefont {Chu},
  \citenamefont {Kharel}, \citenamefont {Yoon}, \citenamefont {Frunzio},
  \citenamefont {Rakich},\ and\ \citenamefont {Schoelkopf}}]{Chu2018}%
  \BibitemOpen
  \bibfield  {author} {\bibinfo {author} {\bibfnamefont {Y.}~\bibnamefont
  {Chu}}, \bibinfo {author} {\bibfnamefont {P.}~\bibnamefont {Kharel}},
  \bibinfo {author} {\bibfnamefont {T.}~\bibnamefont {Yoon}}, \bibinfo {author}
  {\bibfnamefont {L.}~\bibnamefont {Frunzio}}, \bibinfo {author} {\bibfnamefont
  {P.~T.}\ \bibnamefont {Rakich}}, \ and\ \bibinfo {author} {\bibfnamefont
  {R.~J.}\ \bibnamefont {Schoelkopf}},\ }\href@noop {} {\bibfield  {journal}
  {\bibinfo  {journal} {Nature}\ }\textbf {\bibinfo {volume} {563}},\ \bibinfo
  {pages} {666} (\bibinfo {year} {2018})}\BibitemShut {NoStop}%
\bibitem [{\citenamefont {Wollack}\ \emph {et~al.}(2022)\citenamefont
  {Wollack}, \citenamefont {Cleland}, \citenamefont {Gruenke}, \citenamefont
  {Wang}, \citenamefont {Arrangoiz-Arriola},\ and\ \citenamefont
  {Safavi-Naeini}}]{wollack2022}%
  \BibitemOpen
  \bibfield  {author} {\bibinfo {author} {\bibfnamefont {E.~A.}\ \bibnamefont
  {Wollack}}, \bibinfo {author} {\bibfnamefont {A.~Y.}\ \bibnamefont
  {Cleland}}, \bibinfo {author} {\bibfnamefont {R.~G.}\ \bibnamefont
  {Gruenke}}, \bibinfo {author} {\bibfnamefont {Z.}~\bibnamefont {Wang}},
  \bibinfo {author} {\bibfnamefont {P.}~\bibnamefont {Arrangoiz-Arriola}}, \
  and\ \bibinfo {author} {\bibfnamefont {A.~H.}\ \bibnamefont
  {Safavi-Naeini}},\ }\href@noop {} {\bibfield  {journal} {\bibinfo  {journal}
  {Nature}\ }\textbf {\bibinfo {volume} {604}},\ \bibinfo {pages} {463}
  (\bibinfo {year} {2022})}\BibitemShut {NoStop}%
\bibitem [{\citenamefont {Manenti}\ \emph {et~al.}(2017)\citenamefont
  {Manenti}, \citenamefont {Kockum}, \citenamefont {Patterson}, \citenamefont
  {Behrle}, \citenamefont {Rahamim}, \citenamefont {Tancredi}, \citenamefont
  {Nori},\ and\ \citenamefont {Leek}}]{manenti2017}%
  \BibitemOpen
  \bibfield  {author} {\bibinfo {author} {\bibfnamefont {R.}~\bibnamefont
  {Manenti}}, \bibinfo {author} {\bibfnamefont {A.~F.}\ \bibnamefont {Kockum}},
  \bibinfo {author} {\bibfnamefont {A.}~\bibnamefont {Patterson}}, \bibinfo
  {author} {\bibfnamefont {T.}~\bibnamefont {Behrle}}, \bibinfo {author}
  {\bibfnamefont {J.}~\bibnamefont {Rahamim}}, \bibinfo {author} {\bibfnamefont
  {G.}~\bibnamefont {Tancredi}}, \bibinfo {author} {\bibfnamefont
  {F.}~\bibnamefont {Nori}}, \ and\ \bibinfo {author} {\bibfnamefont {P.~J.}\
  \bibnamefont {Leek}},\ }\href@noop {} {\bibfield  {journal} {\bibinfo
  {journal} {Nature communications}\ }\textbf {\bibinfo {volume} {8}},\
  \bibinfo {pages} {1} (\bibinfo {year} {2017})}\BibitemShut {NoStop}%
\bibitem [{\citenamefont {Reed}\ \emph {et~al.}(2017)\citenamefont {Reed},
  \citenamefont {Mayer}, \citenamefont {Teufel}, \citenamefont {Burkhart},
  \citenamefont {Pfaff}, \citenamefont {Reagor}, \citenamefont {Sletten},
  \citenamefont {Ma}, \citenamefont {Schoelkopf}, \citenamefont {Knill} \emph
  {et~al.}}]{reed2017faithful}%
  \BibitemOpen
  \bibfield  {author} {\bibinfo {author} {\bibfnamefont {A.}~\bibnamefont
  {Reed}}, \bibinfo {author} {\bibfnamefont {K.}~\bibnamefont {Mayer}},
  \bibinfo {author} {\bibfnamefont {J.}~\bibnamefont {Teufel}}, \bibinfo
  {author} {\bibfnamefont {L.}~\bibnamefont {Burkhart}}, \bibinfo {author}
  {\bibfnamefont {W.}~\bibnamefont {Pfaff}}, \bibinfo {author} {\bibfnamefont
  {M.}~\bibnamefont {Reagor}}, \bibinfo {author} {\bibfnamefont
  {L.}~\bibnamefont {Sletten}}, \bibinfo {author} {\bibfnamefont
  {X.}~\bibnamefont {Ma}}, \bibinfo {author} {\bibfnamefont {R.}~\bibnamefont
  {Schoelkopf}}, \bibinfo {author} {\bibfnamefont {E.}~\bibnamefont {Knill}},
  \emph {et~al.},\ }\href@noop {} {\bibfield  {journal} {\bibinfo  {journal}
  {Nature Physics}\ }\textbf {\bibinfo {volume} {13}},\ \bibinfo {pages} {1163}
  (\bibinfo {year} {2017})}\BibitemShut {NoStop}%
\bibitem [{\citenamefont {Ghirardi}\ \emph {et~al.}(1990)\citenamefont
  {Ghirardi}, \citenamefont {Pearle},\ and\ \citenamefont
  {Rimini}}]{Ghirardi1990}%
  \BibitemOpen
  \bibfield  {author} {\bibinfo {author} {\bibfnamefont {G.~C.}\ \bibnamefont
  {Ghirardi}}, \bibinfo {author} {\bibfnamefont {P.}~\bibnamefont {Pearle}}, \
  and\ \bibinfo {author} {\bibfnamefont {A.}~\bibnamefont {Rimini}},\ }\href
  {\doibase 10.1103/PhysRevA.42.78} {\bibfield  {journal} {\bibinfo  {journal}
  {Physical Review A}\ }\textbf {\bibinfo {volume} {42}},\ \bibinfo {pages}
  {78} (\bibinfo {year} {1990})}\BibitemShut {NoStop}%
\bibitem [{\citenamefont {Percival}(1994)}]{Percival2006}%
  \BibitemOpen
  \bibfield  {author} {\bibinfo {author} {\bibfnamefont {I.~C.}\ \bibnamefont
  {Percival}},\ }\href@noop {} {\bibfield  {journal} {\bibinfo  {journal}
  {Proceedings of the Royal Society of London. Series A: Mathematical and
  Physical Sciences}\ }\textbf {\bibinfo {volume} {447}},\ \bibinfo {pages}
  {189} (\bibinfo {year} {1994})}\BibitemShut {NoStop}%
\bibitem [{\citenamefont {Manley}\ \emph {et~al.}(2020)\citenamefont {Manley},
  \citenamefont {Wilson}, \citenamefont {Stump}, \citenamefont {Grin},\ and\
  \citenamefont {Singh}}]{manley2020}%
  \BibitemOpen
  \bibfield  {author} {\bibinfo {author} {\bibfnamefont {J.}~\bibnamefont
  {Manley}}, \bibinfo {author} {\bibfnamefont {D.~J.}\ \bibnamefont {Wilson}},
  \bibinfo {author} {\bibfnamefont {R.}~\bibnamefont {Stump}}, \bibinfo
  {author} {\bibfnamefont {D.}~\bibnamefont {Grin}}, \ and\ \bibinfo {author}
  {\bibfnamefont {S.}~\bibnamefont {Singh}},\ }\href@noop {} {\bibfield
  {journal} {\bibinfo  {journal} {Physical review letters}\ }\textbf {\bibinfo
  {volume} {124}},\ \bibinfo {pages} {151301} (\bibinfo {year}
  {2020})}\BibitemShut {NoStop}%
\bibitem [{\citenamefont {Di{\'o}si}(1989)}]{diosi1989models}%
  \BibitemOpen
  \bibfield  {author} {\bibinfo {author} {\bibfnamefont {L.}~\bibnamefont
  {Di{\'o}si}},\ }\href@noop {} {\bibfield  {journal} {\bibinfo  {journal}
  {Physical Review A}\ }\textbf {\bibinfo {volume} {40}},\ \bibinfo {pages}
  {1165} (\bibinfo {year} {1989})}\BibitemShut {NoStop}%
\bibitem [{\citenamefont {Penrose}(2014)}]{penrose2014gravitization}%
  \BibitemOpen
  \bibfield  {author} {\bibinfo {author} {\bibfnamefont {R.}~\bibnamefont
  {Penrose}},\ }\href@noop {} {\bibfield  {journal} {\bibinfo  {journal}
  {Foundations of Physics}\ }\textbf {\bibinfo {volume} {44}},\ \bibinfo
  {pages} {557} (\bibinfo {year} {2014})}\BibitemShut {NoStop}%
\bibitem [{\citenamefont {O’Connell}\ \emph {et~al.}(2010)\citenamefont
  {O’Connell}, \citenamefont {Hofheinz}, \citenamefont {Ansmann},
  \citenamefont {Bialczak}, \citenamefont {Lenander}, \citenamefont {Lucero},
  \citenamefont {Neeley}, \citenamefont {Sank}, \citenamefont {Wang},
  \citenamefont {Weides} \emph {et~al.}}]{o2010quantum}%
  \BibitemOpen
  \bibfield  {author} {\bibinfo {author} {\bibfnamefont {A.~D.}\ \bibnamefont
  {O’Connell}}, \bibinfo {author} {\bibfnamefont {M.}~\bibnamefont
  {Hofheinz}}, \bibinfo {author} {\bibfnamefont {M.}~\bibnamefont {Ansmann}},
  \bibinfo {author} {\bibfnamefont {R.~C.}\ \bibnamefont {Bialczak}}, \bibinfo
  {author} {\bibfnamefont {M.}~\bibnamefont {Lenander}}, \bibinfo {author}
  {\bibfnamefont {E.}~\bibnamefont {Lucero}}, \bibinfo {author} {\bibfnamefont
  {M.}~\bibnamefont {Neeley}}, \bibinfo {author} {\bibfnamefont
  {D.}~\bibnamefont {Sank}}, \bibinfo {author} {\bibfnamefont {H.}~\bibnamefont
  {Wang}}, \bibinfo {author} {\bibfnamefont {M.}~\bibnamefont {Weides}},  \emph
  {et~al.},\ }\href@noop {} {\bibfield  {journal} {\bibinfo  {journal}
  {Nature}\ }\textbf {\bibinfo {volume} {464}},\ \bibinfo {pages} {697}
  (\bibinfo {year} {2010})}\BibitemShut {NoStop}%
\bibitem [{\citenamefont {Palomaki}\ \emph {et~al.}(2013)\citenamefont
  {Palomaki}, \citenamefont {Teufel}, \citenamefont {Simmonds},\ and\
  \citenamefont {Lehnert}}]{palomaki2013entangling}%
  \BibitemOpen
  \bibfield  {author} {\bibinfo {author} {\bibfnamefont {T.}~\bibnamefont
  {Palomaki}}, \bibinfo {author} {\bibfnamefont {J.}~\bibnamefont {Teufel}},
  \bibinfo {author} {\bibfnamefont {R.}~\bibnamefont {Simmonds}}, \ and\
  \bibinfo {author} {\bibfnamefont {K.~W.}\ \bibnamefont {Lehnert}},\
  }\href@noop {} {\bibfield  {journal} {\bibinfo  {journal} {Science}\ }\textbf
  {\bibinfo {volume} {342}},\ \bibinfo {pages} {710} (\bibinfo {year}
  {2013})}\BibitemShut {NoStop}%
\bibitem [{\citenamefont {Aref}\ \emph {et~al.}(2016)\citenamefont {Aref},
  \citenamefont {Delsing}, \citenamefont {Ekstr{\"o}m}, \citenamefont {Kockum},
  \citenamefont {Gustafsson}, \citenamefont {Johansson}, \citenamefont {Leek},
  \citenamefont {Magnusson},\ and\ \citenamefont {Manenti}}]{Aref2015}%
  \BibitemOpen
  \bibfield  {author} {\bibinfo {author} {\bibfnamefont {T.}~\bibnamefont
  {Aref}}, \bibinfo {author} {\bibfnamefont {P.}~\bibnamefont {Delsing}},
  \bibinfo {author} {\bibfnamefont {M.~K.}\ \bibnamefont {Ekstr{\"o}m}},
  \bibinfo {author} {\bibfnamefont {A.~F.}\ \bibnamefont {Kockum}}, \bibinfo
  {author} {\bibfnamefont {M.~V.}\ \bibnamefont {Gustafsson}}, \bibinfo
  {author} {\bibfnamefont {G.}~\bibnamefont {Johansson}}, \bibinfo {author}
  {\bibfnamefont {P.~J.}\ \bibnamefont {Leek}}, \bibinfo {author}
  {\bibfnamefont {E.}~\bibnamefont {Magnusson}}, \ and\ \bibinfo {author}
  {\bibfnamefont {R.}~\bibnamefont {Manenti}},\ }in\ \href@noop {} {\emph
  {\bibinfo {booktitle} {Superconducting devices in quantum optics}}}\
  (\bibinfo  {publisher} {Springer},\ \bibinfo {year} {2016})\ pp.\ \bibinfo
  {pages} {217--244}\BibitemShut {NoStop}%
\bibitem [{\citenamefont {Nielsen}\ \emph {et~al.}(2017)\citenamefont
  {Nielsen}, \citenamefont {Tsaturyan}, \citenamefont {M{\o}ller},
  \citenamefont {Polzik},\ and\ \citenamefont
  {Schliesser}}]{nielsen2017multimode}%
  \BibitemOpen
  \bibfield  {author} {\bibinfo {author} {\bibfnamefont {W.~H.~P.}\
  \bibnamefont {Nielsen}}, \bibinfo {author} {\bibfnamefont {Y.}~\bibnamefont
  {Tsaturyan}}, \bibinfo {author} {\bibfnamefont {C.~B.}\ \bibnamefont
  {M{\o}ller}}, \bibinfo {author} {\bibfnamefont {E.~S.}\ \bibnamefont
  {Polzik}}, \ and\ \bibinfo {author} {\bibfnamefont {A.}~\bibnamefont
  {Schliesser}},\ }\href@noop {} {\bibfield  {journal} {\bibinfo  {journal}
  {Proceedings of the National Academy of Sciences}\ }\textbf {\bibinfo
  {volume} {114}},\ \bibinfo {pages} {62} (\bibinfo {year} {2017})}\BibitemShut
  {NoStop}%
\bibitem [{\citenamefont {Hong}\ \emph {et~al.}(2017)\citenamefont {Hong},
  \citenamefont {Riedinger}, \citenamefont {Marinkovi{\'c}}, \citenamefont
  {Wallucks}, \citenamefont {Hofer}, \citenamefont {Norte}, \citenamefont
  {Aspelmeyer},\ and\ \citenamefont {Gr{\"o}blacher}}]{hong2017hanbury}%
  \BibitemOpen
  \bibfield  {author} {\bibinfo {author} {\bibfnamefont {S.}~\bibnamefont
  {Hong}}, \bibinfo {author} {\bibfnamefont {R.}~\bibnamefont {Riedinger}},
  \bibinfo {author} {\bibfnamefont {I.}~\bibnamefont {Marinkovi{\'c}}},
  \bibinfo {author} {\bibfnamefont {A.}~\bibnamefont {Wallucks}}, \bibinfo
  {author} {\bibfnamefont {S.~G.}\ \bibnamefont {Hofer}}, \bibinfo {author}
  {\bibfnamefont {R.~A.}\ \bibnamefont {Norte}}, \bibinfo {author}
  {\bibfnamefont {M.}~\bibnamefont {Aspelmeyer}}, \ and\ \bibinfo {author}
  {\bibfnamefont {S.}~\bibnamefont {Gr{\"o}blacher}},\ }\href@noop {}
  {\bibfield  {journal} {\bibinfo  {journal} {Science}\ }\textbf {\bibinfo
  {volume} {358}},\ \bibinfo {pages} {203} (\bibinfo {year}
  {2017})}\BibitemShut {NoStop}%
\bibitem [{\citenamefont {Sletten}\ \emph {et~al.}(2019)\citenamefont
  {Sletten}, \citenamefont {Moores}, \citenamefont {Viennot},\ and\
  \citenamefont {Lehnert}}]{sletten2019}%
  \BibitemOpen
  \bibfield  {author} {\bibinfo {author} {\bibfnamefont {L.~R.}\ \bibnamefont
  {Sletten}}, \bibinfo {author} {\bibfnamefont {B.~A.}\ \bibnamefont {Moores}},
  \bibinfo {author} {\bibfnamefont {J.~J.}\ \bibnamefont {Viennot}}, \ and\
  \bibinfo {author} {\bibfnamefont {K.~W.}\ \bibnamefont {Lehnert}},\
  }\href@noop {} {\bibfield  {journal} {\bibinfo  {journal} {Physical Review
  X}\ }\textbf {\bibinfo {volume} {9}},\ \bibinfo {pages} {021056} (\bibinfo
  {year} {2019})}\BibitemShut {NoStop}%
\bibitem [{\citenamefont {Satzinger}\ \emph {et~al.}(2018)\citenamefont
  {Satzinger}, \citenamefont {Zhong}, \citenamefont {Chang}, \citenamefont
  {Peairs}, \citenamefont {Bienfait}, \citenamefont {Chou}, \citenamefont
  {Cleland}, \citenamefont {Conner}, \citenamefont {Dumur}, \citenamefont
  {Grebel} \emph {et~al.}}]{satzinger2018quantum}%
  \BibitemOpen
  \bibfield  {author} {\bibinfo {author} {\bibfnamefont {K.~J.}\ \bibnamefont
  {Satzinger}}, \bibinfo {author} {\bibfnamefont {Y.}~\bibnamefont {Zhong}},
  \bibinfo {author} {\bibfnamefont {H.-S.}\ \bibnamefont {Chang}}, \bibinfo
  {author} {\bibfnamefont {G.~A.}\ \bibnamefont {Peairs}}, \bibinfo {author}
  {\bibfnamefont {A.}~\bibnamefont {Bienfait}}, \bibinfo {author}
  {\bibfnamefont {M.-H.}\ \bibnamefont {Chou}}, \bibinfo {author}
  {\bibfnamefont {A.}~\bibnamefont {Cleland}}, \bibinfo {author} {\bibfnamefont
  {C.~R.}\ \bibnamefont {Conner}}, \bibinfo {author} {\bibfnamefont
  {{\'E}.}~\bibnamefont {Dumur}}, \bibinfo {author} {\bibfnamefont
  {J.}~\bibnamefont {Grebel}},  \emph {et~al.},\ }\href@noop {} {\bibfield
  {journal} {\bibinfo  {journal} {Nature}\ }\textbf {\bibinfo {volume} {563}},\
  \bibinfo {pages} {661} (\bibinfo {year} {2018})}\BibitemShut {NoStop}%
\bibitem [{\citenamefont {Aspelmeyer}\ \emph {et~al.}(2014)\citenamefont
  {Aspelmeyer}, \citenamefont {Kippenberg},\ and\ \citenamefont
  {Marquardt}}]{aspelmeyer2014cavity}%
  \BibitemOpen
  \bibfield  {author} {\bibinfo {author} {\bibfnamefont {M.}~\bibnamefont
  {Aspelmeyer}}, \bibinfo {author} {\bibfnamefont {T.~J.}\ \bibnamefont
  {Kippenberg}}, \ and\ \bibinfo {author} {\bibfnamefont {F.}~\bibnamefont
  {Marquardt}},\ }\href@noop {} {\bibfield  {journal} {\bibinfo  {journal}
  {Reviews of Modern Physics}\ }\textbf {\bibinfo {volume} {86}},\ \bibinfo
  {pages} {1391} (\bibinfo {year} {2014})}\BibitemShut {NoStop}%
\bibitem [{\citenamefont {Mancini}\ \emph {et~al.}(1997)\citenamefont
  {Mancini}, \citenamefont {Man'ko},\ and\ \citenamefont
  {Tombesi}}]{mancini1997}%
  \BibitemOpen
  \bibfield  {author} {\bibinfo {author} {\bibfnamefont {S.}~\bibnamefont
  {Mancini}}, \bibinfo {author} {\bibfnamefont {V.}~\bibnamefont {Man'ko}}, \
  and\ \bibinfo {author} {\bibfnamefont {P.}~\bibnamefont {Tombesi}},\
  }\href@noop {} {\bibfield  {journal} {\bibinfo  {journal} {Physical Review
  A}\ }\textbf {\bibinfo {volume} {55}},\ \bibinfo {pages} {3042} (\bibinfo
  {year} {1997})}\BibitemShut {NoStop}%
\bibitem [{\citenamefont {Bose}\ \emph {et~al.}(1997)\citenamefont {Bose},
  \citenamefont {Jacobs},\ and\ \citenamefont {Knight}}]{Bose1997}%
  \BibitemOpen
  \bibfield  {author} {\bibinfo {author} {\bibfnamefont {S.}~\bibnamefont
  {Bose}}, \bibinfo {author} {\bibfnamefont {K.}~\bibnamefont {Jacobs}}, \ and\
  \bibinfo {author} {\bibfnamefont {P.}~\bibnamefont {Knight}},\ }\href@noop {}
  {\bibfield  {journal} {\bibinfo  {journal} {Physical Review A}\ }\textbf
  {\bibinfo {volume} {56}},\ \bibinfo {pages} {4175} (\bibinfo {year}
  {1997})}\BibitemShut {NoStop}%
\bibitem [{\citenamefont {Kharel}\ \emph {et~al.}(2019)\citenamefont {Kharel},
  \citenamefont {Harris}, \citenamefont {Kittlaus}, \citenamefont {Renninger},
  \citenamefont {Otterstrom}, \citenamefont {Harris},\ and\ \citenamefont
  {Rakich}}]{kharel2019high}%
  \BibitemOpen
  \bibfield  {author} {\bibinfo {author} {\bibfnamefont {P.}~\bibnamefont
  {Kharel}}, \bibinfo {author} {\bibfnamefont {G.~I.}\ \bibnamefont {Harris}},
  \bibinfo {author} {\bibfnamefont {E.~A.}\ \bibnamefont {Kittlaus}}, \bibinfo
  {author} {\bibfnamefont {W.~H.}\ \bibnamefont {Renninger}}, \bibinfo {author}
  {\bibfnamefont {N.~T.}\ \bibnamefont {Otterstrom}}, \bibinfo {author}
  {\bibfnamefont {J.~G.}\ \bibnamefont {Harris}}, \ and\ \bibinfo {author}
  {\bibfnamefont {P.~T.}\ \bibnamefont {Rakich}},\ }\href@noop {} {\bibfield
  {journal} {\bibinfo  {journal} {Science advances}\ }\textbf {\bibinfo
  {volume} {5}},\ \bibinfo {pages} {eaav0582} (\bibinfo {year}
  {2019})}\BibitemShut {NoStop}%
\bibitem [{\citenamefont {Rakich}\ \emph {et~al.}(2012)\citenamefont {Rakich},
  \citenamefont {Reinke}, \citenamefont {Camacho}, \citenamefont {Davids},\
  and\ \citenamefont {Wang}}]{Rakich2012}%
  \BibitemOpen
  \bibfield  {author} {\bibinfo {author} {\bibfnamefont {P.~T.}\ \bibnamefont
  {Rakich}}, \bibinfo {author} {\bibfnamefont {C.}~\bibnamefont {Reinke}},
  \bibinfo {author} {\bibfnamefont {R.}~\bibnamefont {Camacho}}, \bibinfo
  {author} {\bibfnamefont {P.}~\bibnamefont {Davids}}, \ and\ \bibinfo {author}
  {\bibfnamefont {Z.}~\bibnamefont {Wang}},\ }\href
  {http://link.aps.org/doi/10.1103/PhysRevX.2.011008} {\bibfield  {journal}
  {\bibinfo  {journal} {Physical Review X}\ }\textbf {\bibinfo {volume} {2}},\
  \bibinfo {pages} {11008} (\bibinfo {year} {2012})}\BibitemShut {NoStop}%
\bibitem [{\citenamefont {Kharel}\ \emph {et~al.}(2016)\citenamefont {Kharel},
  \citenamefont {Behunin}, \citenamefont {Renninger},\ and\ \citenamefont
  {Rakich}}]{Kharel2016}%
  \BibitemOpen
  \bibfield  {author} {\bibinfo {author} {\bibfnamefont {P.}~\bibnamefont
  {Kharel}}, \bibinfo {author} {\bibfnamefont {R.~O.}\ \bibnamefont {Behunin}},
  \bibinfo {author} {\bibfnamefont {W.~H.}\ \bibnamefont {Renninger}}, \ and\
  \bibinfo {author} {\bibfnamefont {P.~T.}\ \bibnamefont {Rakich}},\
  }\href@noop {} {\bibfield  {journal} {\bibinfo  {journal} {Physical Review
  A}\ }\textbf {\bibinfo {volume} {93}},\ \bibinfo {pages} {063806} (\bibinfo
  {year} {2016})}\BibitemShut {NoStop}%
\bibitem [{\citenamefont {Jaynes}\ and\ \citenamefont
  {Cummings}(1963)}]{jaynes1963comparison}%
  \BibitemOpen
  \bibfield  {author} {\bibinfo {author} {\bibfnamefont {E.~T.}\ \bibnamefont
  {Jaynes}}\ and\ \bibinfo {author} {\bibfnamefont {F.~W.}\ \bibnamefont
  {Cummings}},\ }\href@noop {} {\bibfield  {journal} {\bibinfo  {journal}
  {Proceedings of the IEEE}\ }\textbf {\bibinfo {volume} {51}},\ \bibinfo
  {pages} {89} (\bibinfo {year} {1963})}\BibitemShut {NoStop}%
\bibitem [{\citenamefont {Arfken}\ and\ \citenamefont
  {Weber}(2005)}]{arfken1999mathematical}%
  \BibitemOpen
  \bibfield  {author} {\bibinfo {author} {\bibfnamefont {G.~B.}\ \bibnamefont
  {Arfken}}\ and\ \bibinfo {author} {\bibfnamefont {H.~J.}\ \bibnamefont
  {Weber}},\ }\href@noop {} {\emph {\bibinfo {title} {Mathematical methods for
  physicists}}}\ (\bibinfo  {publisher} {Elsevier Academic Press; 6th
  edition},\ \bibinfo {year} {2005})\BibitemShut {NoStop}%
\bibitem [{\citenamefont {Shore}\ and\ \citenamefont
  {Knight}(1993)}]{shore1993jaynes}%
  \BibitemOpen
  \bibfield  {author} {\bibinfo {author} {\bibfnamefont {B.~W.}\ \bibnamefont
  {Shore}}\ and\ \bibinfo {author} {\bibfnamefont {P.~L.}\ \bibnamefont
  {Knight}},\ }\href@noop {} {\bibfield  {journal} {\bibinfo  {journal}
  {Journal of Modern Optics}\ }\textbf {\bibinfo {volume} {40}},\ \bibinfo
  {pages} {1195} (\bibinfo {year} {1993})}\BibitemShut {NoStop}%
\bibitem [{\citenamefont {Johansson}\ \emph {et~al.}(2012)\citenamefont
  {Johansson}, \citenamefont {Nation},\ and\ \citenamefont
  {Nori}}]{johansson2012qutip}%
  \BibitemOpen
  \bibfield  {author} {\bibinfo {author} {\bibfnamefont {J.~R.}\ \bibnamefont
  {Johansson}}, \bibinfo {author} {\bibfnamefont {P.~D.}\ \bibnamefont
  {Nation}}, \ and\ \bibinfo {author} {\bibfnamefont {F.}~\bibnamefont
  {Nori}},\ }\href@noop {} {\bibfield  {journal} {\bibinfo  {journal} {Computer
  Physics Communications}\ }\textbf {\bibinfo {volume} {183}},\ \bibinfo
  {pages} {1760} (\bibinfo {year} {2012})}\BibitemShut {NoStop}%
\bibitem [{\citenamefont {Chan}\ \emph {et~al.}(2009)\citenamefont {Chan},
  \citenamefont {Eichenfield}, \citenamefont {Camacho},\ and\ \citenamefont
  {Painter}}]{Chan2009}%
  \BibitemOpen
  \bibfield  {author} {\bibinfo {author} {\bibfnamefont {J.}~\bibnamefont
  {Chan}}, \bibinfo {author} {\bibfnamefont {M.}~\bibnamefont {Eichenfield}},
  \bibinfo {author} {\bibfnamefont {R.}~\bibnamefont {Camacho}}, \ and\
  \bibinfo {author} {\bibfnamefont {O.}~\bibnamefont {Painter}},\ }\href@noop
  {} {\bibfield  {journal} {\bibinfo  {journal} {Optics Express}\ }\textbf
  {\bibinfo {volume} {17}},\ \bibinfo {pages} {3802} (\bibinfo {year}
  {2009})}\BibitemShut {NoStop}%
\bibitem [{\citenamefont {Gerry}\ \emph {et~al.}(2005)\citenamefont {Gerry},
  \citenamefont {Knight},\ and\ \citenamefont
  {Knight}}]{gerry2005introductory}%
  \BibitemOpen
  \bibfield  {author} {\bibinfo {author} {\bibfnamefont {C.}~\bibnamefont
  {Gerry}}, \bibinfo {author} {\bibfnamefont {P.}~\bibnamefont {Knight}}, \
  and\ \bibinfo {author} {\bibfnamefont {P.~L.}\ \bibnamefont {Knight}},\
  }\href@noop {} {\emph {\bibinfo {title} {Introductory quantum optics}}}\
  (\bibinfo  {publisher} {Cambridge university press},\ \bibinfo {year}
  {2005})\BibitemShut {NoStop}%
  \bibitem{Schwinger1952}
J. Schwinger, Unpublished report, NYO-3071 (1952).
  
\end{thebibliography}

%

\end{document}